# Magnetic Field Induced Quantum Metric Dipole in Dirac Semimetal $Cd_3As_2$

Tong-Yang Zhao[1], An-Qi Wang[1], Zhen-Tao Zhang[1], Zheng-Yang Cao[1], Xing-Yu Liu[1] and Zhi-Min Liao[1,2]*

[1]State Key Laboratory for Mesoscopic Physics and Frontiers Science Center for Nano-optoelectronics, School of Physics, Peking University, Beijing 100871, China.

[2]Hefei National Laboratory, Hefei 230088, China.

*Email: liaozm@pku.edu.cn

The quantum geometry, comprising Berry curvature and quantum metric, plays a fundamental role in governing electron transport phenomena in solids. Recent studies show that the quantum metric dipole drives scattering-free nonlinear Hall effect in topological antiferromagnets, prompting the questions of whether this effect can occur in nonmagnetic systems and be externally tuned by a magnetic field. Our work addresses these frontiers by demonstrating that the quantum metric dipole is actively tuned by an external magnetic field to generate a time-reversal-odd nonlinear Hall response in a nonmagnetic topological Dirac semimetal $Cd_3As_2$. Alongside the well-known chiral-anomaly-induced negative longitudinal magnetoresistance, an exotic nonlinear planar Hall effect emerges with increasing magnetic field. Careful scaling analysis indicates that this nonlinear planar Hall effect is controlled by the magnetic-field-modulated quantum metric dipole. Constructing a $k \cdot p$ effective model of the Dirac bands under Zeeman and orbital coupling, we derive the evolution of the quantum metric dipole as a function of the magnetic field, providing a comprehensive explanation of the experimental results. Our results establish a band-structure-based strategy for engineering nonlinear magnetotransport in nonmagnetic materials via the quantum metric dipole, opening a pathway toward magnetic-field–tunable nonlinear quantum devices.

The quantum geometric tensor, which encapsulates both the Berry curvature and quantum metric, has been instrumental in advancing the understanding of diverse electronic phenomena in condensed matter systems [1,2]. The Berry curvature can induce transversal charge transport, governing phenomena such as the anomalous Hall effect [3-10], orbital magnetization [11-14], and valley Hall effect [15,16]. On the other hand, the quantum metric, a measure of the distance between quantum states in Hilbert space [17,18], plays an essential role in the physics of fractional Chern insulators, flat-band systems, and other emergent states of matter [17,19-23]. Recently, the quantum metric dipole (QMD), a $k$-space dipole moment of the quantum metric, has been identified as a driving mechanism behind the scattering-independent nonlinear Hall effect. This effect has been observed in pristine $MnBi_2Te_4$ [24] and $MnBi_2Te_4$/black phosphorus heterostructures [25], as well as in antiferromagnetic $Mn_3Sn$/Pt systems where the nonlinear Hall signal persists at room temperature [26]. However, because it requires broken time-reversal symmetry for QMD to emerge, its manifestation has so far been confined to materials with simultaneous magnetism [27,28], severely limiting its applicability to nonmagnetic materials. A pivotal question arises: Can QMD-induced nonlinear transport emerge in nonmagnetic systems and be externally controlled?

To address this, we propose a paradigm shift: engineering QMD in nonmagnetic Dirac semimetals via external magnetic fields. The Dirac semimetal $Cd_3As_2$—a prototypical system with hidden quantum geometry structure—provides an ideal platform. Pristine $Cd_3As_2$ features symmetry protected Dirac cones [29-37] and exhibits pronounced quantum geometric effects, manifested in a nontrivial Berry phase via quantum oscillations [29-32] and quantum Hall effect [33,34]. Crucially, applying a magnetic field splits its Dirac nodes into Weyl pairs, augmenting the QMD via $k$-space separation of Weyl points [38]. This tunable character coincides with chiral anomaly effects [39-42], enabling unprecedented control over QMD-driven nonlinear transport. Furthermore, recent theoretical studies predict that an external magnetic field parallel to the bias current can lead to a nonlinear Hall response, known as the nonlinear planar Hall effect (NPHE) [43-47], which includes a QMD mechanism from the magnetic-field-perturbed band structure [43,44,47]. Experimental demonstration of this exotic transport phenomenon remains highly desirable.

In this work, we show that an external magnetic field alone can induce and modulate the QMD in $Cd_3As_2$, giving rise to a robust, intrinsic NPHE that persists up to room temperature. The aforementioned scenario of magnetic-field-induced Dirac band splitting is responsible for the QMD generation, and the linear magnetic-field dependence of nonlinear Hall signals is crucially determined by the $k$-space separation of Weyl nodes likewise. Our results not only extend QMD-driven nonlinear magnetotransport to nonmagnetic materials but also suggest that magnetic-field–tunable quantum geometry may underlie a broad array of exotic transport phenomena in quantum materials.

We carried out magnetotransport measurements in $Cd_3As_2$ nanoplates grown by chemical vapor deposition technique (see Supplemental Note 1 [48] for details). The nanoplates possess (112) surface plane and $[1\bar{1}0]$ edge direction of body-centered tetragonal structure with space group $I4_1/acd$, as confirmed by previous transmission electron microscope results [67]. Although an ideal crystal lattice preserves inversion symmetry, strain—inevitably introduced via thermal-expansion mismatch or microfabrication processes—breaks that symmetry.

We first establish the signatures of chiral anomaly and second-order nonlinear Hall effect in our samples at $T = 2$ K, under an in-plane magnetic field applied parallel to the nanoplate edge [Fig. 1(a)]. An AC bias $I_{ac}$ at a fixed frequency of 17.777 Hz is applied to the sample, with both longitudinal and transversal responses at fundamental and second harmonics recorded. The first-order longitudinal signal exhibits clear signatures of chiral anomaly-induced negative magnetoresistance (NMR) [Fig. 1(b); see also Supplemental Note 3] [39,41,42,48], along with a weak anti-localization (WAL) effect (Fig. S1 [48]). The NMR is most pronounced at a gate voltage of $V_g \approx$ -10 V, where the Fermi level approaches the Dirac point, as indicated by the resistance peak in the transfer curve [Fig. 1(c)]. In this regime, the nonlinear Hall response shows only a weak dependence on $V_g$ (Supplemental Note 4 [48]).

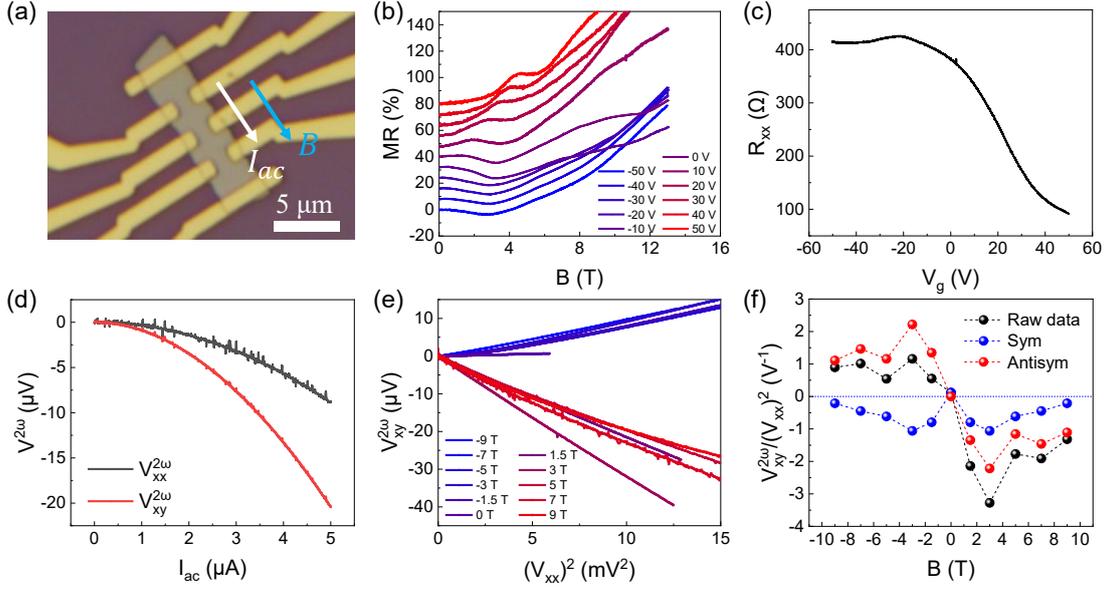

**FIG. 1. Chiral anomaly and nonlinear Hall effect in Cd$_3$As$_2$ at $T = 2$ K.**

(a) Optical image of the Cd$_3$As$_2$ nanoplate device. Arrows indicate the applied parallel current and magnetic field.

(b) Longitudinal magnetoresistance (MR), defined as $[R_{xx}(B) - R_{xx}(0)]/R_{xx}(0) \times 100\%$, under various gate voltages. The curves, which should originate at MR = 0, are vertically shifted for clearer visualization.

(c) Transfer curve of the Cd$_3$As$_2$ nanoplate device.

(d) Second-order longitudinal ($V_{xx}^{2\omega}$, black) and Hall ($V_{xy}^{2\omega}$, red) signals with respect to AC bias at $V_g = -10$ V and $B = 1$ T.

(e) $V_{xy}^{2\omega}$ versus the square of longitudinal voltage $(V_{xx})^2$, under $V_g = -10$ V and various magnetic fields.

(f) Nonlinear Hall ratio $V_{xy}^{2\omega}/(V_{xx})^2$: raw data, symmetric, and antisymmetric components plotted as a function of magnetic field $B$.

Figure 1(d) plots together the second-order longitudinal ($V_{xx}^{2\omega}$) and Hall ($V_{xy}^{2\omega}$) responses measured at $V_g = -10$ V and $B = 1$ T. The prominent $V_{xy}^{2\omega}$ clearly demonstrate the domination of Hall response. Figure 1(e) shows the nonlinear Hall signal $V_{xy}^{2\omega}$ versus the quadratic of longitudinal voltage $V_{xx}$, across various magnetic fields, at the Dirac point $V_g = -10$ V. The $V_{xy}^{2\omega} - (V_{xx})^2$ curves exhibit clear linearity. After symmetrizing $V_{xx}$ (Supplemental Note 1 [48]), we define the nonlinear Hall generation ratio as $V_{xy}^{2\omega}/(V_{xx})^2$, and summarize its dependence on magnetic field $B$ in Fig. 1(f).

Our primary focus is on the time-reversal-odd nonlinear Hall signal. Nevertheless, the nonlinear Hall ratio in Fig. 1(f) shows a deviation from perfect antisymmetry with respect to magnetic field, i. e. $V_{xy}^{2\omega}/(V_{xx})^2|_{+B} \neq -V_{xy}^{2\omega}/(V_{xx})^2|_{-B}$. The $B$-symmetric and antisymmetric components of $V_{xy}^{2\omega}/(V_{xx})^2$ are presented by blue and red circles in Fig. 1(f), respectively. Under relatively low magnetic field $B$, both components show roughly linear dependence on $B$, reaching their respective maxima at $B = 3$ T. Beyond such turning point, both signals saturate and then slightly decrease with increasing $B$ (see continuously measured $V_{xy}^{2\omega}/(V_{xx})^2 - B$ relation in Supplemental Note 5 [48] for better illustration of these features). Notably, such turning point in nonlinear Hall ratio aligns with the turning point of NMR in the linear longitudinal response [Fig. 1(b)]. Apart from the $B$-antisymmetric signal of our interest, the $B$-symmetric signal can also provide complementary evidence towards the incorporated quantum geometry features. We employ scaling analysis to the $B$-antisymmetric and symmetric parts individually for a more comprehensive illustration of the underlying physical scenario (see Supplemental Note 6 [48] for analysis of the $B$-symmetric signal).

Figure 2(a) shows the $B$-odd nonlinear Hall ratio $\frac{E_{xy,\text{odd}}^{2\omega}}{(E_{xx})^2}$ under various magnetic fields and temperatures, with the Fermi level tuned close to the Dirac point (see data under extra temperature points within 10 ~ 70 K in Supplemental Note 7 [48]). Here $E_{xy,\text{odd}}^{2\omega} = V_{xy,\text{odd}}^{2\omega}/W$ and $E_{xx} = V_{xx}/L$ are the corresponding electric fields, with $V_{xy,\text{odd}}^{2\omega} \equiv [V_{xy}^{2\omega}(B) - V_{xy}^{2\omega}(-B)]/2$ the $B$-odd part of nonlinear Hall signal; $W$ and $L$ are the width and length of the conduction channel, respectively. Remarkably, the nonlinear Hall signal persists up to 270 K (limited by the temperature stability of our measurement system), with a unprecedent strength of 133.6 μV under 1 mA bias and moderate 1 T magnetic field. Figure 2(b) plots together the nonlinear Hall ratio $\frac{E^{2\omega}_{xy,\text{odd}}}{(E_{xx})^2}$ and the longitudinal conductivity $\sigma_{xx}$ with respect to temperature, grouped by magnetic field strength. For scaling analysis, we perform parabolic fitting (Supplemental Note 8 [48]) to the nonlinear Hall conductivity $\sigma^{(o)}_{yxx} \equiv \frac{j^{2\omega}_y}{E^2_{xx}} \approx \frac{\sigma_{xx} E^{2\omega}_{xy,\text{odd}}}{E^2_{xx}}$ as a function of $\frac{\sigma_{xx}}{\sigma_0}$, where $\sigma_0$ is the reference longitudinal conductivity at $T = 2$ K and the superscript $(o)$ denotes time-reversal-odd. All experimental data can be well-fitted by the parabolic dependence (see Supplemental Note 8.2 [48] for detailed discussion about the determination of the proper scaling formula)

$$\sigma^{(o)}_{yxx} = A_0 + A_1 \left(\frac{\sigma_{xx}}{\sigma_0}\right) + A_2 \left(\frac{\sigma_{xx}}{\sigma_0}\right)^2. \qquad (1)$$

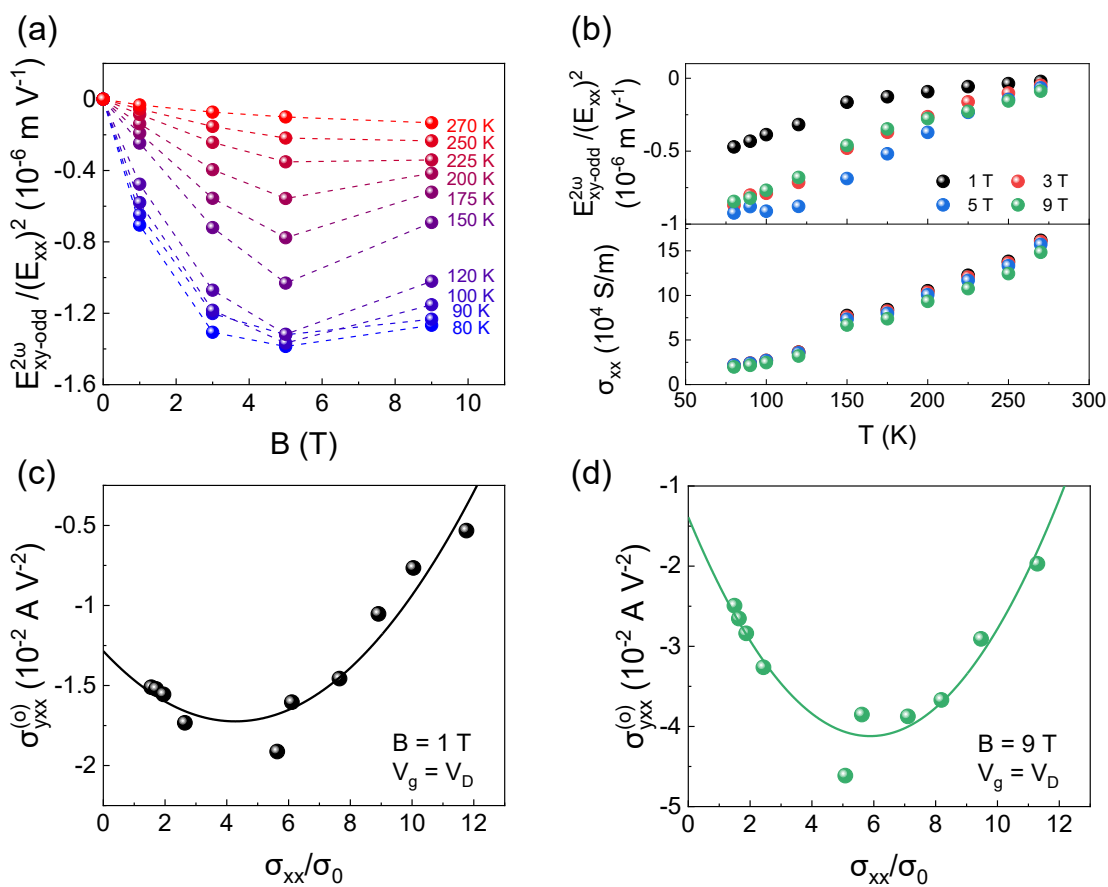

**FIG. 2. Scaling analysis of the time-reversal-odd nonlinear Hall signal.**

(a) The $\frac{E^{2\omega}_{xy-\text{odd}}}{(E_{xx})^2} - B$ relation measured at various temperatures, with Fermi level close to the Dirac point.

(b) Nonlinear Hall ratio $\frac{E^{2\omega}_{xy-odd}}{(E_{xx})^2}$ and longitudinal conductivity $\sigma_{xx}$ as functions of temperature.

(c), (d) Nonlinear Hall conductivity $\sigma^{(o)}_{yxx}$ plotted against $\frac{\sigma_{xx}}{\sigma_0}$ under (c) $B = 1\,\text{T}$ and (d) $B = 9\,\text{T}$ as nonlinear Hall scaling relations. The solid curves are parabolic fittings to corresponding data.

Regarding the physical implications of these fitting parameters, recent experimental studies suggest the direct association between the $\tau$-independent zeroth-order term ($A_0$) and the band-intrinsic contribution [24-26,44,68]. The remaining $\tau$-dependent terms all correspond to time-reversal-odd extrinsic mechanisms. Figures 2(c) and 2(d) show the time-reversal-odd scaling results under $B = 1\,\text{T}$ and $B = 9\,\text{T}$, respectively. Under moderate magnetic field ($B = 1\,\text{T}$), the intercept of parabolic fitting curve as $\frac{\sigma_{xx}}{\sigma_0} \to 0$ is comparable to the overall amplitude of $\sigma^{(o)}_{yxx}$, indicating the dominance of $A_0$ term. In contrast, at a strong magnetic field ($B = 9\,\text{T}$), the intercept constitutes only a small fraction of the maximum value of $\sigma^{(o)}_{yxx}$, suggesting that extrinsic scattering becomes the primary contributor.

Recent researches on $B$-free intrinsic nonlinear Hall effect [59,68] and $B$-dependent NPHE [44] emphasize the role of QMD-induced correction to transport behaviors. In these scenarios, the QMD enters the charge transport by generating an anomalous transversal velocity in carriers (Supplemental Note 2 [48]), similar to the Berry curvature and leads to macroscopic Hall effect under broken inversion and time-reversal symmetry. By supposing broken inversion symmetry due to strain during sample growth and fabrication, and with magnetic field further breaking the time-reversal symmetry, the QMD nonlinear Hall response naturally becomes feasible. Furthermore, the high magnetic-sensitivity of DSM band structure [37,38] renders strong tunability of emergent QMD (Fig. 3), providing a consistent explanation to our results as we demonstrate in the following.

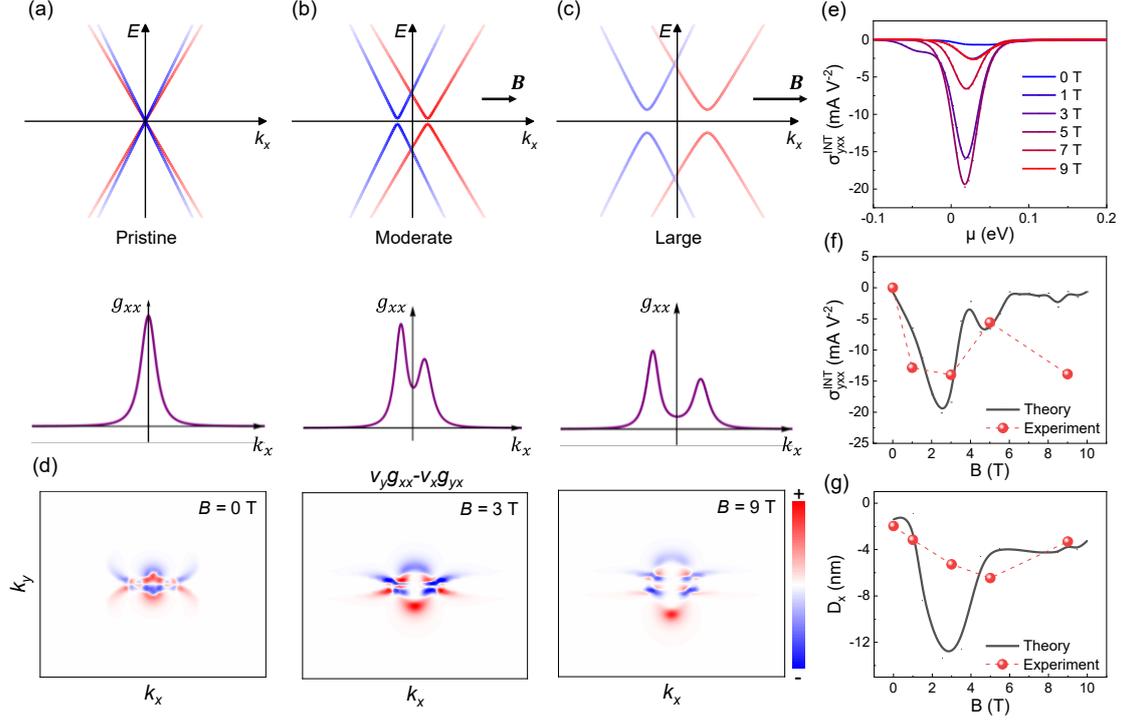

**FIG. 3. Quantum geometry evolution under magnetic field.**

(a)-(c) Schematic band structure of DSM with broken inversion symmetry, under (a) zero magnetic field, (b) weak field and (c) strong field. Blue and red colors refer to Weyl cones with opposite chirality, and thickness of the colors indicates the magnitude of associated Berry curvature. Lower panels in (a)-(c): schematic distribution of quantum metric component $g_{xx}$ along the field direction $k_x$.

(d) Numerically calculated $k$-space distribution of $v_y g_{xx} - v_x g_{yx}$ with different magnetic fields.

(e) Calculated intrinsic nonlinear Hall conductivity $\sigma_{yxx}^{INT}$ as a function of chemical potential $\mu$ under various magnetic fields.

(f), (g) Comparison between theoretical and experimental estimation of (f) $\sigma_{yxx}^{INT}$ and (g) Berry curvature dipole $D_x$ near the Dirac point.

Figures 3(a)-(c) depict the schematic evolution of one Dirac cone under external magnetic field applied along a low-symmetry direction. The inversion-symmetry-broken DSM is considered, manifested by different group velocities for Weyl cones with opposite chirality [60]. In the absence of magnetic field [Fig. 3(a)], the two Weyl

cones overlap in the momentum space without a bandgap. Although quantum metric components are expected to show divergent behavior in gapless band, following $g_{\alpha\beta}(k) \propto (\Delta E)^{-2}$ with $g_{\alpha\beta}(k)$ the quantum metric component (Supplemental Note 2 [48], band index $n$ is omitted) and $\Delta E$ the energy gap, the time reversal symmetry guarantees a symmetric distribution as $g_{\alpha\beta}(k) = g_{\alpha\beta}(-k)$. Consequently, the net QMD vanishes, as illustrated in the lower panel of Fig. 3(a).

When magnetic field is applied, Weyl cones with opposite chirality become separated along the magnetic field direction, and finite gaps emerge due to the reduction of $C_4$ lattice symmetry concurrently [Fig. 3(b)(c)]. The broken time-reversal symmetry permits a finite QMD, yet its magnitude depends on two opposing trends: the separated Weyl cones form a more extended dipole structure, while formation of bandgap suppresses the quantum metric globally. When magnetic field remains small, the induced band gap is negligible, qualitatively preserving the Weyl cone dispersion, which aligns with the chiral anomaly scenario [39] [Fig. 3(b)]. The nearly gapless band structure continues to have quantum metric concentrated near the band edges, leading to an overall enhancing trend in QMD, answering for the observed increase in the nonlinear Hall signal under low magnetic fields. When magnetic field further increases towards the breakdown of Weyl cone dispersion with a notable gap [Fig. 3(c)], the overall reduction of quantum metric value constrains the QMD from further enhanced, so QMD eventually saturates at high magnetic fields, consistent with our experimental observations in such regime.

The above illustration of QMD evolution captures the essential physics of generating and manipulating QMD structure in non-magnetic DSM system. For further quantitative confirmation of our proposed scenario, we carry out numerical calculations of quantum geometry basing on an effective model of $Cd_3As_2$ (see Supplemental Note 11 [48]). Following conclusion from previous study [25], the intrinsic nonlinear Hall conductivity in a multi-band system can be decomposed into two fractions: one arising from QMD ($D_{QM} \equiv \int_k (v_y g_{xx} - v_x g_{yx}) \delta(\varepsilon - \varepsilon_F)$) representing contribution from the nearest band in energy, and the other known as additional intraband contributions (AIC) that accounts all the other bands. Our calculations show that the quantity $(v_y g_{xx} - v_x g_{yx})$ becomes highlighted at band edges and crossings, and its evolution driven by magnetic field is essentially determined by the underlying band structure (Fig. S9) [48]. Figure 3(d) shows the structure of $(v_y g_{xx} - v_x g_{yx})$ in the transport plane under various magnetic fields applied along $k_x$. The evolution of such quantity aligns with our expectation from the simplified model: the magnetic field leads to a more delocalized and asymmetric distribution of quantum metric, while also suppressing its overall magnitude.

Figure 3(e) shows the calculated intrinsic nonlinear Hall conductivity $\sigma_{yxx}^{\text{INT}}$ as a function of chemical potential $\mu$, exhibiting its strong band-edge-concentrating feature. Figure 3(f) compares the numerical and experimental results (Fig. 2) for $\sigma_{yxx}^{\text{INT}}$, the latter of which are taken as the $\tau$-independent contribution from the overall experimentally obtained nonlinear Hall conductivity. The calculation quantitatively matches the experiments and reproduces the anticipated nonmonotonic dependence on $B$. However, under large magnetic field ($B = 9$ T), a clear deviation between theory and experiment emerges. This discrepancy can likely be attributed to the theoretically proposed zeroth-order extrinsic (ZOE) contributions in $\tau^0$ term of $\sigma_{yxx}^{(o)}$, arising from side-jump, skew scattering, or collaboration of these scatterings [57,58], coexisting with QMD contribution. At $B = 9$ T, it is plausible that the majority of $\tau^0$ term arises from these ZOE contributions, whereas intrinsic contribution is indeed much smaller than $A_0$ term (see Supplemental Note 8 [48] for further discussions). Nevertheless, in the low-field regime, our study provides strong evidence for the dominance of the QMD-driven intrinsic nonlinear Hall transport, manifesting accessibility of QMD-related physics in non-magnetic topological systems. For time-reversal-even nonlinear Hall response, we perform similar theoretical analysis and comparison with experiments (Fig. 3(g), see also Supplemental Notes 6 and 11 [48]). The coherent results further indicate the magnetic tunability of quantum geometry realized by band evolution in DSM.

While the self-consistent theoretical and experimental analysis above properly identify the quantum geometric effects, it is important to remain cautious about trivial interferers including thermal effects, circuit capacitive coupling, contact diode effect, and electrode misalignment, which could influence the nonlinear signal measurements and complicate data interpretation. To address this, we perform additional transport measurements and analysis to evaluate these side-contributions. Our results indicate that these factors have a negligible impact on the nonlinear Hall signal driven by the band-intrinsic quantum geometry (see Supplemental Note 12 [48]).

In conclusion, our study unveils the magnetic field-induced and tunable QMD structure in the non-magnetic DSM $Cd_3As_2$, probed through time-reversal-odd nonlinear Hall signals. The strong nonlinear response and its tunability can be attributed to the magnetic field-driven Weyl cone separation, which is a band-structure effect that can be easily generalized in broader context of quantum materials. The robustness of the signal against thermal fluctuation suggests potential applicability in functional devices. This highlights the significance of exploiting quantum metric physics in a broader range of non-magnetic topological materials, with the methods developed in this study being directly applicable to future investigations. For instance, topological semimetals such as $WTe_2$ and $TaIrTe_4$, as well as moiré systems with flat bands, are poised to leverage remarkable quantum metric effects under external field modulation. Furthermore, we recognize that tuning QMD with a magnetic field offers an effective approach to probing topological phase transitions. The QMD-driven nonlinear Hall response can reveal the closing and opening of topological gaps, as well as band degeneracy lifting, providing a clearer view of quantum criticality in systems with rich phase diagrams [69-71].

**Acknowledgements**

This work was supported by the National Natural Science Foundation of China (Grant Nos. 62425401 and 62321004), and Innovation Program for Quantum Science and Technology (Grant No. 2021ZD0302403).

# Supplemental Material for

# Magnetic Field Induced Quantum Metric Dipole in Dirac Semimetal Cd$_3$As$_2$

Tong-Yang Zhao[1], An-Qi Wang[1], Zhen-Tao Zhang[1], Zheng-Yang Cao[1], Xing-Yu Liu[1] and Zhi-Min Liao[1,2]*

[1]State Key Laboratory for Mesoscopic Physics and Frontiers Science Center for Nano-optoelectronics, School of Physics, Peking University, Beijing 100871, China.
[2]Hefei National Laboratory, Hefei 230088, China.
*Email: liaozm@pku.edu.cn

**Table of contents:**

**Note 1. Device fabrication, transport measurement setup and nonlinear Hall data processing**

Nanoplate structures of $Cd_3As_2$ were grown by chemical vapor deposition (CVD) method. Raw $Cd_3As_2$ powder with stoichiometric ratio was heated to 750 ℃ to decompose, then cooled down on silicon substrate to form single crystal with low impurity concentration. Nanoplates with ideal size ($2-3$ μm in width and $>10$ μm in length) and thickness (less than 100 nm) were selected and transferred onto $Si/SiO_2$ substrates using glass tips. Polymethyl methacrylate (PMMA) was employed as resist to define the Hall bar electrode pattern on the samples, after standard electron beam lithography and development procedures. Before depositing Ti (5 nm) / Au (150 nm) electrodes, we performed argon plasma etching to remove the natural oxide layer of the samples and ensure ohmic contact.

The electrical transport experiments were conducted in an Oxford commercial cryostat with a base temperature of 2 K. AC bias supply and first-, second-harmonic signal detection were performed by Stanford Research Systems SR830 lock-in amplifiers, with base frequency at 17.777 Hz unless stated otherwise. DC bias was generated by Keithley 2400 SourceMeter, serving as gate voltage of the device through the 285 nm $SiO_2$ dielectric layer of the substrate.

During electrical transport measurements, electrodes misalignment often leads to mixing of longitudinal and Hall signals. Concerning the nonlinear Hall ratio $V_{xy}^{2\omega}/(V_{xx})^2$, the longitudinal voltage $V_{xx}$ should first receive a symmetrization, $V_{xx} \to [V_{xx}(+B) + V_{xx}(-B)]/2$, so that the evaluation about the parity of nonlinear Hall signal under time reversal is not undermined. In Fig. 1(e) in the main text, we also have chosen to show the $V_{xy}^{2\omega} - V_{xx}^2$ relations instead of $V_{xy}^{2\omega} - I_{ac}^2$ relations for the same reason, since the latter case cannot remove the influence of mismatch and asymmetric $V_{xx}(B)$ in the following scaling analysis. By plotting $V_{xy}^{2\omega}$ versus the square of symmetrized $V_{xx}$, the slopes of the curves can better resemble the nonlinear Hall conductivity with less systematic error. For nonlinear Hall signal $V_{xy}^{2\omega}$, we perform anti-symmetrization $V^{2\omega}_{xy-\text{odd}} \equiv [V^{2\omega}_{xy}(+B) - V^{2\omega}_{xy}(-B)]/2$ to obtain the time-reversal-odd part, and symmetrization $V^{2\omega}_{xy-\text{even}} \equiv [V^{2\omega}_{xy}(+B) + V^{2\omega}_{xy}(-B)]/2$ for the time-reversal-even part.

**Note 2. Symmetry of quantum geometry and its associated nonlinear Hall transport**

The quantum geometric tensor is defined as $T_{\alpha\beta}^n = \sum_{m\neq n} \mathcal{A}_{nm}^\alpha \mathcal{A}_{mn}^\beta$, with $\mathcal{A}_{nm}^\alpha \equiv \langle u_n | i\partial_{k_\alpha} | u_m \rangle$ the interband Berry connection and $|u_n\rangle$ the periodic part of Bloch wavefunction. The quantum metric $g$ and Berry curvature $\Omega$ are linked to the real and imaginary parts of $T$, as $g_{\alpha\beta}^n = \text{Re}[T_{\alpha\beta}^n]$ and $\Omega_{\alpha\beta}^n = -2\text{Im}[T_{\alpha\beta}^n]$. When system possesses inversion symmetry, the constraint $T_{\alpha\beta}^n(+\boldsymbol{k}) = T_{\alpha\beta}^n(-\boldsymbol{k})$ is expected, leading to $g_{\alpha\beta}^n(+\boldsymbol{k}) = g_{\alpha\beta}^n(-\boldsymbol{k})$ and $\Omega_{\alpha\beta}^n(+\boldsymbol{k}) = \Omega_{\alpha\beta}^n(-\boldsymbol{k})$. Such symmetric distribution in $k$-space causes the dipole moment $\int_k \partial_{k_\gamma} g_{\alpha\beta}^n$ or $\int_k \partial_{k_\gamma} \Omega_{\alpha\beta}^n$ to vanish on any direction $\gamma$. On the other hand, when system possesses time reversal symmetry, the constraint becomes $T_{\alpha\beta}^n(+\boldsymbol{k}) = T_{\alpha\beta}^n(-\boldsymbol{k})^*$, leading to $g_{\alpha\beta}^n(+\boldsymbol{k}) = g_{\alpha\beta}^n(-\boldsymbol{k})$ and $\Omega_{\alpha\beta}^n(+\boldsymbol{k}) = -\Omega_{\alpha\beta}^n(-\boldsymbol{k})$. As a result, the QMD still vanishes, yet the BCD survives under an antisymmetric distribution. In addition, for a PT-symmetric case, we have $T_{\alpha\beta}^n(+\boldsymbol{k}) = T_{\alpha\beta}^n(+\boldsymbol{k})^*$, which does not restrict the value of quantum metric or its dipole, but forces Berry curvature to vanish anywhere in $k$-space.

The scenario of transversal anomalous velocity driven by Berry curvature $\Omega$ has been well-accepted, following $v_{\text{anom}}^{(1)} \propto E \times \Omega$, and leading to the intrinsic linear anomalous Hall effect [3,4]. The anomalous velocity induced by quantum metric $g$ follows a similar scenario. A bias electric field can perturbatively induce a correction to the Berry connection of the system, $\mathcal{A}^{(1)} = \overleftrightarrow{G}E$, known as the Berry connection polarizability (BCP) mechanism [5,49-51], with

$$G_{\alpha\beta}^n = 2e\text{Re} \sum_{m\neq n} \frac{\mathcal{A}_{nm}^\alpha \mathcal{A}_{mn}^\beta}{\varepsilon_n - \varepsilon_m}$$

the BCP tensor elements. The similar mathematical forms of $G_{\alpha\beta}^n$ and $g_{\alpha\beta}^n = \text{Re}\sum_{m\neq n} \mathcal{A}_{nm}^\alpha \mathcal{A}_{mn}^\beta$ conveys that BCP can be understood as a consequence of quantum metric. Such understanding is rational, considering the fact that quantum metric and Berry curvature mathematically forms a Kramers-Krönig analytic pair. Thus, anomalous velocity induced by quantum metric can be roughly written as $v_{anom}^{(2)} \propto E \times \Omega^{(1)} \sim E \times \nabla \times (gE)$, which scales quadratically with electric field and leads to second-order Hall effect. In the presence of either inversion symmetry or time reversal symmetry, a symmetric distribution $g_{\alpha\beta}^n(k) = g_{\alpha\beta}^n(-k)$ is expected, forcing the net anomalous transport to vanish after summing $v_{anom}^{(2)}$ over all occupied states. Therefore, we require the external magnetic field to break the time reversal symmetry, enabling the emergence of time-reversal-odd nonlinear transport.

**Note 3. Additional experimental evidence and analysis of chiral anomaly**

Previous studies have reported the chiral anomaly effect in the Dirac semimetal $Cd_3As_2$ [39,41,42]. Our magnetotransport data show similar results.

Figure S1 plots the zoom-in details of Fig. 1(b), showing the low-magnetic field MR within $-0.6\,\text{T} < B < 0.6\,\text{T}$, under gate voltages $-50\,\text{V} < V_g < 0\,\text{V}$. A low-field positive MR is observed, which corresponds to the weak anti-localization (WAL) mechanism [52-54].

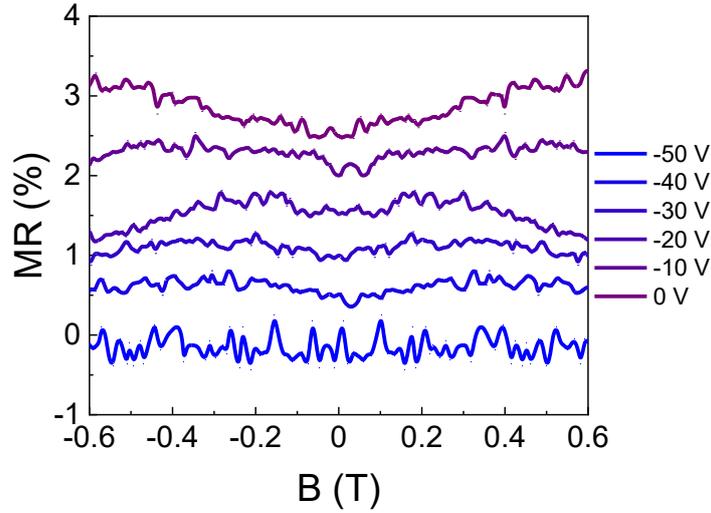

**FIG. S1. Low-field magnetoresistance measured under various gate voltages, as a zoom-in of Fig. 1(b) in the main text.**

Figure S2(a) shows the NMR result under various temperatures, with Fermi level tuned close to the Dirac point (in practice, we interpret the position of $R_{xx}$ maximum in transfer curves as the Dirac point). Clear NMR signature can be observed over a wide range of temperature from 2 to 100 K, and attenuates at 120 K or higher. Our results exhibit much similarity with previous NMR results reported in $Cd_3As_2$ nanowires and nanoplates [39].

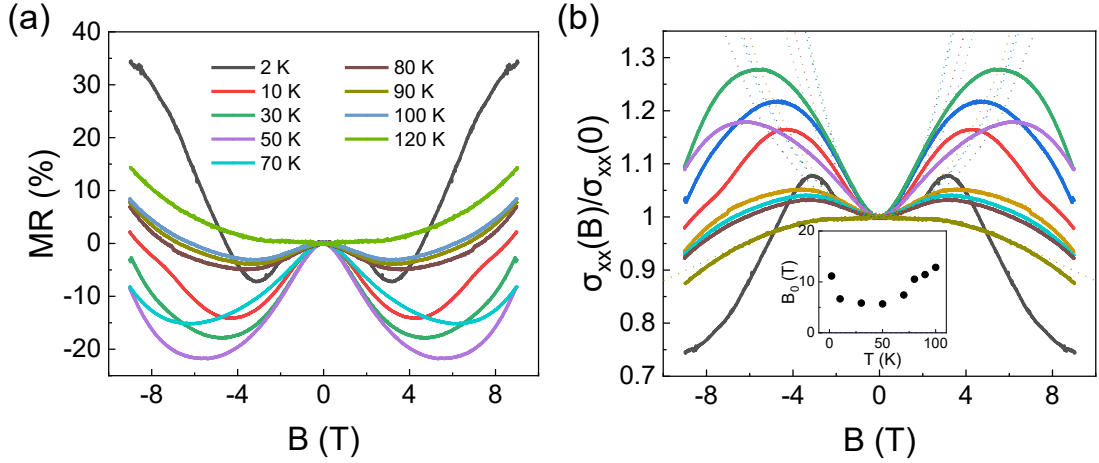

**FIG. S2. Negative magnetoresistance measurements under various temperatures.**
(a) MR data at Dirac point under different temperatures from 2 to 120 K.
(b) Normalized conductivity $\frac{\sigma_{xx}(B)}{\sigma_{xx}(0)}$ as a function of magnetic field, under the same temperatures in (a). Dotted lines are parabolic fittings. Inset: temperature dependence of parameter $B_0$ in equation (S1).

Furthermore, we carry out a simple quantitative analysis by converting MR data to $\sigma_{xx} - B$ relations, as shown in Fig. S2(b). Low-field conductivity is expected to follow a parabolic dependence on $B$ as [42]

$$\frac{\sigma_{xx}(B)}{\sigma_{xx}(0)} = 1 + \frac{B^2}{B_0^2}. \tag{S1}$$

We apply parabolic fitting to the data within the range $[-2\,\mathrm{T}, 2\,\mathrm{T}]$, and get favorable fitting results. The parameter $B_0$ depicts information about band structure and scattering processes. Inset of Fig. S2(b) summarizes $B_0$ as a function of temperature. The values all fall within the reasonable range as reported in previous literatures [41,42], further indicating that the chiral charge transport is responsible for the NMR effect here.

**Note 4. Gate voltage dependence of the time-reversal-odd nonlinear Hall signal**

We further check the gate voltage dependence of the time-reversal-odd nonlinear Hall signal under $T = 2$ K. The signals measured under various gate voltages around $V_g = -10$ V are summarized in Fig. S3. It is clearly observed that these $V_{xy}^{2\omega}/V_{xx}^2 - B$ relations all share the similar feature we have discussed in the main text, namely $V_{xy}^{2\omega}/V_{xx}^2$ linearly depends on $B$ in the low field regime, while saturates when $B$ further increases, and the turning point $B = 3$ T is also universal to all these gate voltages. Therefore, despite of the slight difference in absolute magnitude of the signal, the gate voltage does not prominently affect the behavior of the nonlinear Hall response, as long as the Fermi level remain not too far away from the Dirac point (in our case, the broad window $[-30 \text{ V}, 30 \text{ V}]$ is feasible). In such context, although NMR and WAL results only roughly estimate the position of Dirac point $V_g \approx -10$ V, the influence caused by such inaccuracy can be safely neglected, without violating our major claims.

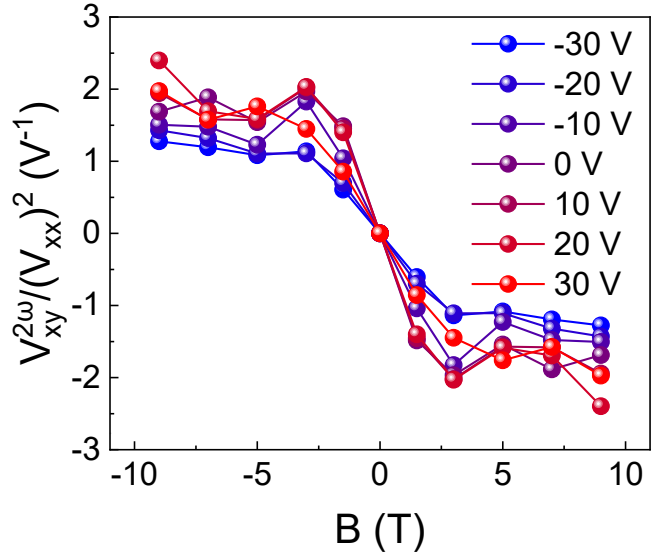

**FIG. S3. Gate voltage dependence of $B$-odd nonlinear Hall signal around the Dirac point $V_g = -10$ V.**

**Note 5. Continuous magnetic field dependence of nonlinear Hall signal**

The continuous relation between the nonlinear Hall signal ratio $V_{xy}^{2\omega}/(V_{xx})^2$ and magnetic field $B$ is shown in Fig. S4, measured with temperature $T = 2$ K and gate voltage $V_g = -10$ V. The curves resemble the discrete data shown in Fig. 1(f), with the typical nonmonotonic feature as discussed in the main text. The magnetic field dependence of corresponding nonlinear Hall conductivity $\sigma_{yxx}$ shown in Fig. S4(b) qualitatively matches the numerical results (Fig. 3(f), see also Supplemental Note 11). The certain quantitative misalignment is due to that the data have not undergone scaling analysis, and represent the sum of all possible contributions besides of intrinsic QMD effects.

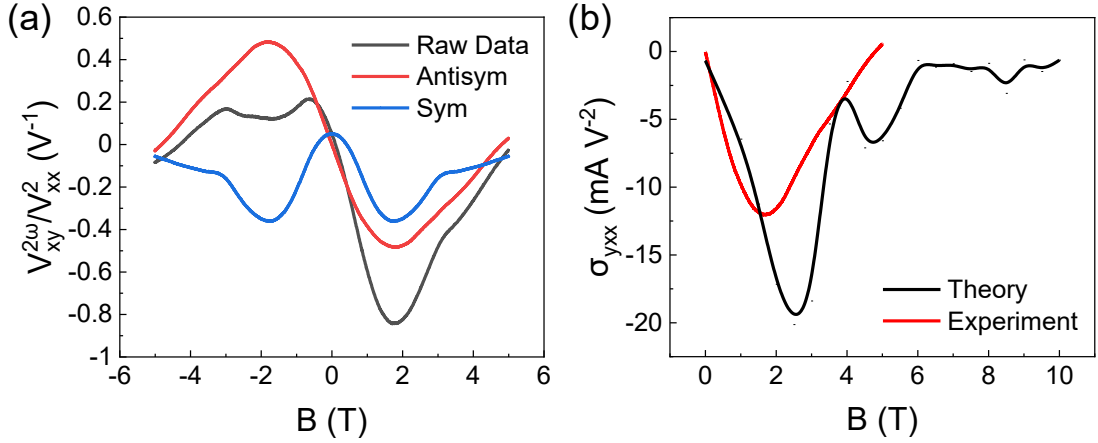

**FIG. S4. Continuously swept $V_{xy}^{2\omega}/V_{xx}^2 - B$ relation under $T = 2$ K.** (a) The raw data (black) is decomposed into symmetric (blue) and antisymmetric (red) parts. (b) The antisymmetric part compared to the numerical result.

**Note 6. Scaling analysis of the time-reversal-even nonlinear Hall effect**

The time-reversal-even part of the nonlinear Hall signal $E^{2\omega}_{xy-\text{even}} \equiv [V^{2\omega}_{xy}(B) + V^{2\omega}_{xy}(-B)]/2W$ under various magnetic fields and temperatures is presented in Fig. S5(a). Scaling law of time-reversal-even signal can be generalized from the nonlinear anomalous Hall scaling (Supplemental Note 8). Therefore, the proper scaling law reads [7,55]

$$\frac{E^{2\omega}_{xy-\text{even}}}{E^2_{xx}} = C_0 + C_1 \left(\frac{\sigma_{xx}}{\sigma_0}\right) + C_2 \left(\frac{\sigma_{xx}}{\sigma_0}\right)^2. \tag{S2}$$

Notice that $\frac{E^{2\omega}_{xy-\text{even}}}{E^2_{xx}}$ rather than the corresponding conductivity $\sigma^{(e)}_{yxx} = \sigma_{xx} \frac{E^{2\omega}_{xy-\text{even}}}{E^2_{xx}}$ is adopted for scaling, because $\tau^0$ term is expected to vanish in $\sigma^{(e)}_{yxx}$. Figure S5(b) provides a representative scaling result of $B$-even nonlinear Hall signal at $B = 5$ T, demonstrating good parabolic fitting. The magnetic field dependence of parameters $C_i$'s is depicted in Fig. S5(c).

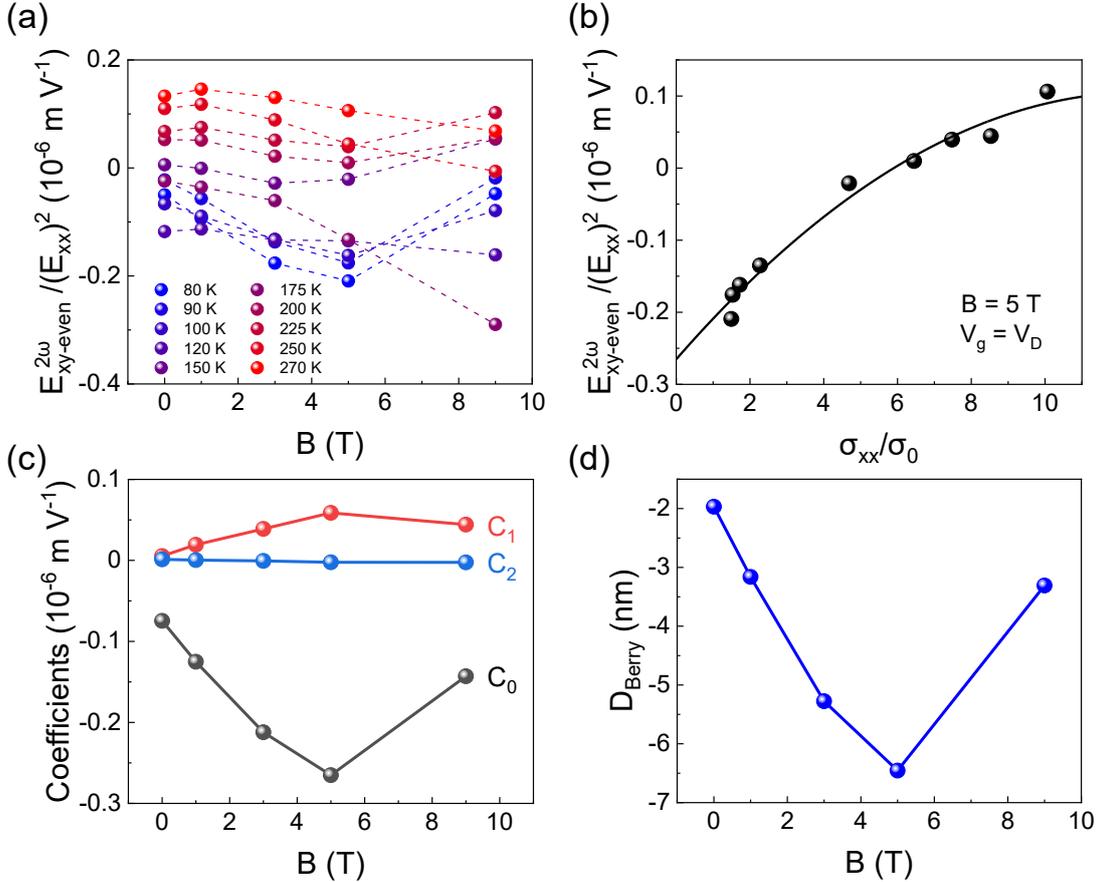

**FIG. S5. Scaling analysis of the time-reversal-even nonlinear Hall signal.**

(a) The $\frac{E^{2\omega}_{xy-\text{even}}}{(E_{xx})^2} - B$ relation measured at various temperatures, with Fermi level close to the Dirac point.

(b) $\frac{E^{2\omega}_{xy-\text{even}}}{(E_{xx})^2}$ versus $\frac{\sigma_{xx}}{\sigma_0}$ as the nonlinear Hall scaling relation under $B = 5$ T.

(c) Magnetic field dependence of the scaling coefficients $C_0$, $C_1$ and $C_2$.

(d) The estimated value of effective BCD $D_{\text{Berry}}$ versus the magnetic field.

Analogous to the nonlinear anomalous Hall scaling, each parameter represents certain composition of variant carrier depletion mechanisms including BCD, side-jump or skew scattering. The high carrier mobility obtained from transfer curve (Supplemental Note 10) suggests the relatively clean, defection-free condition of our sample, effectively ruling out the side-jump and non-Gaussian scattering contributions. This leads to a simplified scaling model (Supplemental Note 8) and the BCD contribution can be determined as $C_{\text{BCD}} = C_0 + \frac{1}{2}C_1$ [55,56]. The effective value of BCD, $D_{\text{Berry}}$, is deduced from the relation [6]

$$D_{\text{Berry}} = \frac{2\hbar^2 n}{m^* e} C_{\text{BCD}} \tag{S3}$$

with $\hbar$ the reduced Planck constant, $e$ the elementary charge, $n$ the carrier density, and $m^* = 0.04 m_e$ the effective mass. The magnetic field dependence of $D_{\text{Berry}}$ is summarized in Fig. S5(d).

**Note 7. Nonlinear Hall data under extra temperatures**

Figure S6(a) shows the magnetic field-dependent nonlinear Hall data within the temperature range 2 ~ 80 K, with Fermi level tuned close to the Dirac point under each temperature. All these data show the presence of a turning point at $B = 3\,\text{T}$ as discussed in the main text, exhibiting consistency across this temperature regime. Nevertheless, these data were not adopted in the scaling analysis due to reasons as follows. It is noteworthy that the resistance-temperature relation $R_{xx} - T$ of Dirac semimetal Cd$_3$As$_2$ is usually accompanied with a kink signature at relatively low temperature [39,41,42], which is also observed in our device (Fig. S6(c)). The metallic $R_{xx} - T$ signature at low temperature $T \leq 70\,\text{K}$ indicates that on top of the general thermal activation, the Cd$_3$As$_2$ device simultaneously experiences a rapid shift in the equilibrium Fermi level, as confirmed by the transfer curves in Fig. S6(d). Even if the nonlinear Hall data in this temperature range are all measured under gate voltages where $R_{xx}$ reaches the respective maximum, the difference in gate voltage could still lead to different patterns of screening effect, interfering with subsequent data interpretation. On the other hand, the Fermi level shifting behavior becomes alleviated under higher temperatures, as shown in Fig. S6(d) as well as the retained semiconductor-like $R_{xx} - T$ behavior in Fig. S6(c) with $T \geq 80\,\text{K}$. Therefore, with gate voltage tuned to a relatively fixed value close to the Dirac point, the high temperature dataset is self-contained to support the valid scaling analysis in the main text.

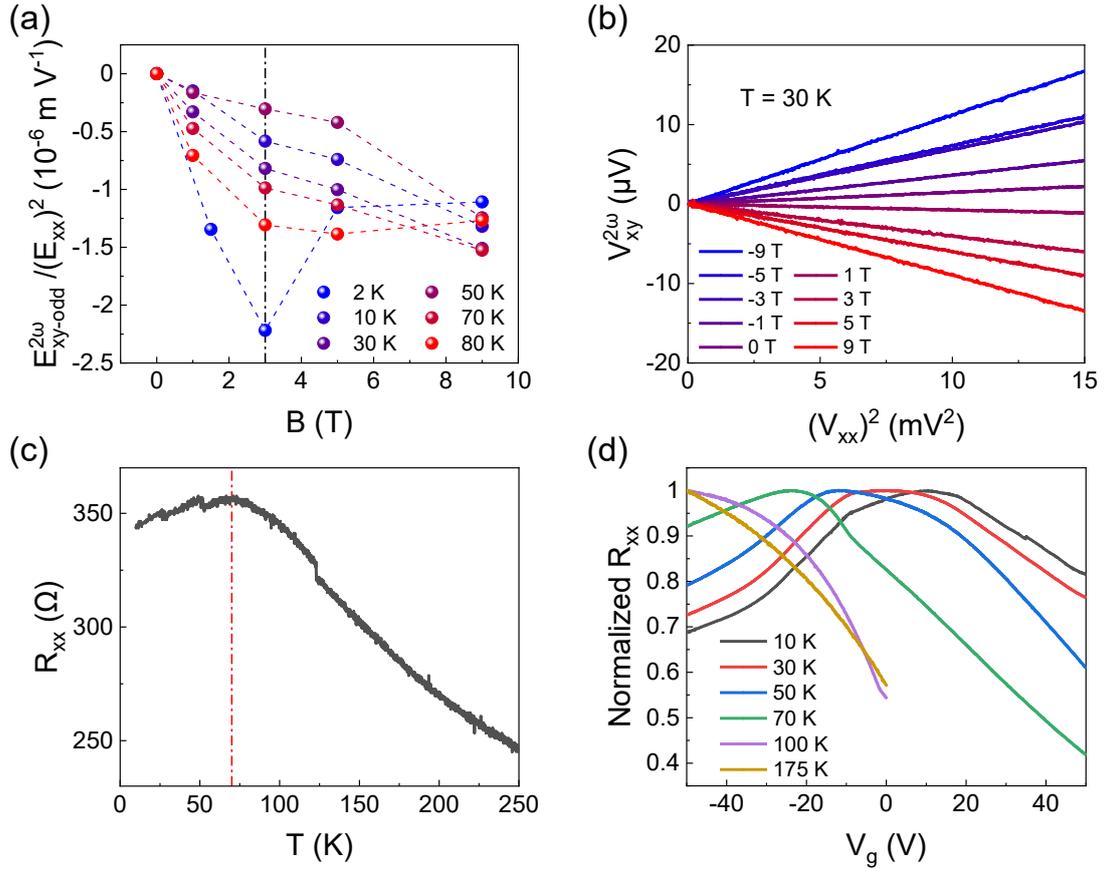

**FIG. S6. Signatures of nonlinear Hall signal and transfer curve under low temperatures.**

(a) The $\frac{E^{2\omega}_{xy-odd}}{(E_{xx})^2} - B$ relation measured under several temperatures from 2 to 80 K, with Fermi level close to the Dirac point. Black dash dot line corresponds to $B = 3$ T.

(b) Nonlinear Hall response $V^{2\omega}_{xy} - (V_{xx})^2$ measured under various magnetic fields and the typical temperature $T = 30$ K.

(c) Temperature dependence of longitudinal resistance $R_{xx} - T$. Red dash dot line corresponds to $T = 70$ K.

(d) Normalized transfer curves $R_{xx} - V_g$ measured under various temperatures from 10 to 175 K.

**Note 8. Rationality about the choices of scaling function for time-reversal-odd and even nonlinear Hall signals**

*Note 8. 1. Scaling law for time-reversal-even nonlinear Hall effect*

We first discuss the proper scaling law for the time-reversal-even nonlinear Hall effect, which is associated with BCD contribution and other extrinsic scattering mechanisms. The second-order nonlinear Hall mechanism can be expressed as BCD deflection, skew scattering, side-jump or their certain combinations [57]. Since all these mechanisms are already present in the zero-field case, the onset of magnetic field should merely give correction to the strength of each mechanism, leaving the overall physical scenario as well as their $\tau$-dependence unaffected.

For the case of nonlinear anomalous Hall effect, the scaling relation reads [7]

$$\frac{E_{xy}^{2\omega}}{(E_{xx})^2} = C_0 + C_1 \left(\frac{\sigma_{xx}}{\sigma_0}\right) + C_2 \left(\frac{\sigma_{xx}}{\sigma_0}\right)^2 + C_2' \left(\frac{\sigma_{xx}^2}{\sigma_0}\right). \tag{S4}$$

Here, $C_0 = C_{\text{BCD}} + C_1^{sj} + C_{11}^{sk,1}$, $C_1 = C_0^{sj} - C_1^{sj} + C_{01}^{sk,1} - 2C_{11}^{sk,1}$, $C_2 = -C_0^{sj} - C_1^{sj} + C_{00}^{sk,1} + C_{11}^{sk,1} - 2C_{01}^{sk,1}$, and $C_2' = C^{sk,2}$ can in principle be obtained from parabolic fitting of experimental data. $C_{\text{BCD}}$ resembles the band-intrinsic contribution from BCD, $C_i^{sj}$ refers to side-jump scattering, and $C_{ij}^{sk,1}$, $C^{sk,2}$ refer to gaussian and non-gaussian skew scattering, respectively; $i(j) = 0$ or $1$ corresponds to scattering with static (impurity) or dynamic (phonon) source, respectively. Basing on previous statements, we can directly apply equation (S4) to fit our time-reversal-even nonlinear Hall data. Better still, following our previous analysis about relative significance of scattering mechanisms in pristine Cd3As2 [56], we can safely exclude side-jump, non-gaussian skew scattering, and multi-source skew scattering in our sample (guaranteed by high device mobility, see Supplemental Note 10). A simplified version of fitting parameters then reads $C_0 = C_{\text{BCD}} + C_{11}^{sk,1}$, $C_1 = -2C_{11}^{sk,1}$, $C_2 = C_{00}^{sk,1} + C_{11}^{sk,1}$ and $C_2' = 0$. This leads to the scaling relation equation (S2), as well as the relation $C_{\text{BCD}} = C_0 + \frac{1}{2}C_1$ for the experimental estimation of BCD [Fig. S5(d)].

*Note 8. 2. Scaling law for time-reversal-odd nonlinear Hall effect*

We then turn to the relatively more complicated time-reversal-odd case. As stated in the main text, a paradigm for practical scaling analysis is still under identification. The time-reversal-odd nature indicates that any corresponding mechanism must be absent in the conventional scenario of nonlinear anomalous Hall effect in non-magnetic materials, making equation (S4) totally irreferable here. We notice that recent theoretical works propose a polynomial scaling law [57,58], including terms up to $\tau^4$, i. e. the general scaling relation reads

$$\sigma_{yxx}^{(o)} = A_0 + A_1 \left(\frac{\sigma_{xx}}{\sigma_0}\right) + A_2 \left(\frac{\sigma_{xx}}{\sigma_0}\right)^2 + A_3 \left(\frac{\sigma_{xx}}{\sigma_0}\right)^3 + A_4 \left(\frac{\sigma_{xx}}{\sigma_0}\right)^4. \quad (S5)$$

Notice that the left-hand-side is the time-reversal-odd nonlinear Hall conductivity $\sigma_{yxx}^{(o)}$ here. All the coefficients $A_i$ are still certain combination of various scattering mechanisms, yet not consisted in the same way as the aforementioned $C_i$'s. Notably, it is claimed that sizable $\tau^3$ and $\tau^4$ terms require the cooperation of multiple scattering sources. In accordance with our analysis about the time-reversal-even scattering (where we ignored the multi-source skew scattering), as well as avoiding over-fitting from too many parameters, we drop the higher-order $\tau$ terms and reduce the relation to a parabolic dependence, and arrive at equation (1) in the main text.

Although the form of scaling has become simpler, the detailed physical implications of these $A_i$'s are still unclear, unlike the time-reversal-even case. However, it is evident that any contribution related to quantum metric dipole must be included in the $\tau^0$ term $A_0$, as theoretical works declare [28,44,51,57-59]. Apart from quantum metric contribution, the rest part of $A_0$ comes from extrinsic scattering, known as zeroth-order extrinsic (ZOE) contributions. Also, the $\tau$-dependent terms $A_1$ and $A_2$ exclusively consist of extrinsic mechanisms. Thus, at present stage, merely a rough estimation about the contribution from quantum metric is possible from experimental data, among the overall nonlinear Hall conductivity. When the value of $A_0$ dominates the entire conductivity $\sigma_{yxx}^{(o)}$, which is the low magnetic field case [Fig. 2(c)], we assert the system is in a weak-extrinsic regime, and $A_0$ mostly resemble the intrinsic contribution. Once $A_0$ weights only a small proportion of $\sigma_{yxx}^{(o)}$ [for instance, compare $\sigma_{yxx}^{(o)}$ at $\frac{\sigma_{xx}}{\sigma_0} \approx 6$, and the intercept at $\frac{\sigma_{xx}}{\sigma_0} = 0$ in Fig. 2(d), where $B$ is large], the system enters a strong-extrinsic regime, and $A_0$ should be dominated by ZOE terms, undermining significance of quantum metric. Such scenario is able to explain the results in Fig. 3(f), where experimental value of $\sigma_{yxx}^{\text{INT}}$ deviates from theoretical one under large magnetic field, indicating an overestimation from $A_0$.

**Note 9. Determination of the sample thickness**

Figure S7 shows the thickness of the $Cd_3As_2$ nanoplate sample in our study, determined from atomic force microscopy (AFM) measurement. The line-cut is assigned in the conduction channel as indicated in the optical image. The thickness is relatively smooth and averages to approximately 95 nm.

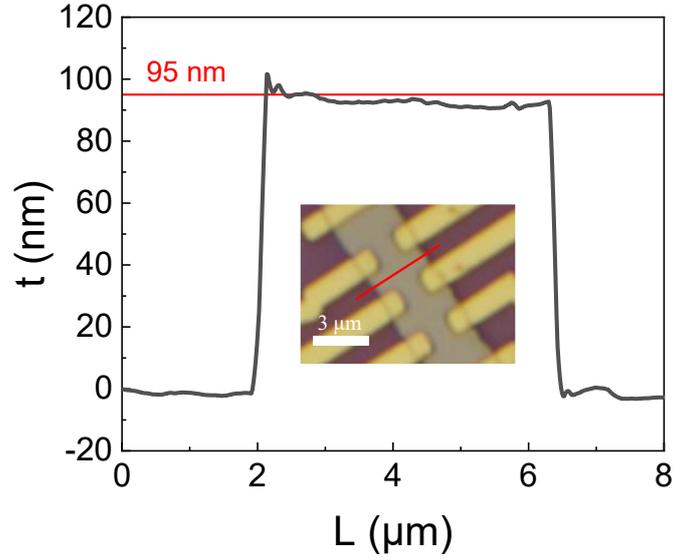

**FIG. S7. AFM profile of the sample conduction channel.** Inset shows the position of the line-cut.

**Note 10. Carrier mobility obtained from transfer curve**

Figure S8 shows the $\sigma_{xx} - V_g$ relation obtained from transfer curve measurement. The electron mobility can be calculated from the formula

$$\mu_e = \frac{\partial \sigma_{xx}}{\partial V_g} \frac{dt}{\varepsilon_0 \varepsilon_{SiO_2}} \tag{S6}$$

with $d = 285$ nm the thickness of SiO$_2$ dielectric layer, $t = 95$ nm the thickness of device, $\varepsilon_0$ the vacuum dielectric constant and $\varepsilon_{SiO_2} = 3.9$ the relative dielectric constant of SiO$_2$. The electron mobility is then estimated as $2.13 \times 10^4$ cm$^2$/(V·s). The electron density in the vicinity of Dirac point is estimated from $n = \frac{\sigma_{xx}}{e\mu}$ to be $8.98 \times 10^{16}$ cm$^{-3}$. Such relatively low carrier concentration guarantees that our device fully exhibits physics close to Dirac point, adding fidelity to the chiral anomaly regime and corresponding scenario of Weyl cone separation under magnetic field.

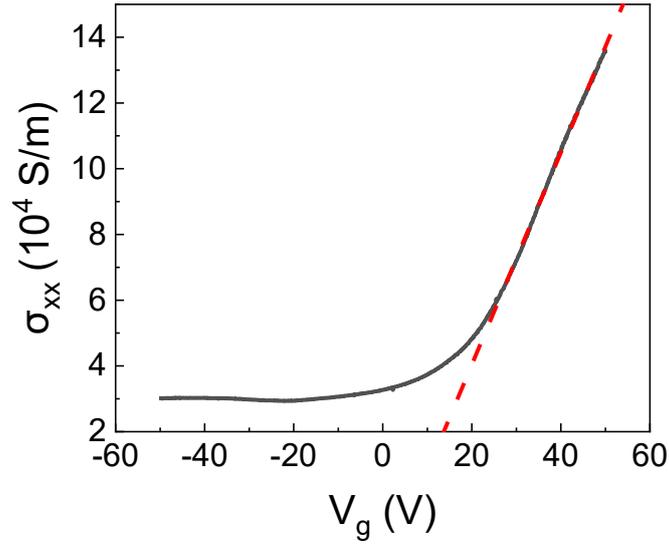

**FIG. S8. Estimation of electron mobility basing on the transfer curve.**

**Note 11. Theoretical modeling and calculations of quantum geometric quantities**

*Note 11. 1. Basic setup of model Hamiltonian with magnetic field*

Several primary setup and methods in our calculation are inherited from Ref. [37]. We use the 4-band effective $k \cdot p$ Hamiltonian that describes the low energy behavior of $Cd_3As_2$ tetragonal structure. The basic Hamiltonian is given as [37,60]

$$\mathcal{H}_0(k) = \epsilon(k) + \begin{pmatrix} M(k) & Dk_- & A^*(k) & B^*(k) \\ Dk_+ & M(k) & B(k) & -A(k) \\ A(k) & B^*(k) & -M(k) & 0 \\ B(k) & -A^*(k) & 0 & -M(k) \end{pmatrix}, \quad (S7)$$

where $\epsilon(k) = E_0 + E_1 k_3^2 + E_2(k_1^2 + k_2^2)$, $M(k) = M_0 + M_1 k_3^2 + M_2(k_1^2 + k_2^2)$, $A(k) = Ak_- + O(k^3)$, and $B(k) = O(k^3)$, $k_\pm = k_1 \pm ik_2$. $k_1, k_2$ and $k_3$ refer to wavevectors along $a, b, c$ crystal axes. The $O(k^3)$ terms recover the trigonal warping signatures of $Cd_3As_2$ with higher Fermi energy, and we omit them in our calculation as we only focus on low-energy chiral anomaly regime. It is noteworthy that the $Dk_\pm$ term is necessary in our model, for it explicitly breaks the inversion symmetry and enables finite dipole structures of quantum geometric quantities.

We apply non-perturbative approach to investigate the effect of magnetic field. This requires modification to the basic Hamiltonian $\mathcal{H}_0$. On the one hand, there is Zeeman coupling term due to interaction between magnetic field and spin, orbital moments: [61]

$$\mathcal{H}_{\text{Zeeman}}(B) = \mu_B \begin{pmatrix} g_{1z}B_3 & g_{1p}B_- & 0 & 0 \\ g_{1p}B_+ & -g_{1z}B_3 & 0 & 0 \\ 0 & 0 & g_{2z}B_3 & g_{2p}B_- \\ 0 & 0 & g_{2p}B_+ & -g_{2z}B_3 \end{pmatrix}, \quad (S8)$$

where $B_\pm = B_1 \pm iB_2$. On the other hand, the gauge field will modify the original momentum into canonical momentum through Peierl's substitution: $\mathcal{H}_0(k) \to \mathcal{H}_0(k - eA)$, with $A$ the vector potential. The full Hamiltonian then reads $\mathcal{H}(k, B) = \mathcal{H}_0(k - eA) + \mathcal{H}_{\text{Zeeman}}(B)$. For model parameters, we use the same values as reported in previous literatures [37,62], summarized in Table S1. The value of $D$ depends on the degree of inversion-breaking of real system, which arises from inherent lattice asymmetry, as well as strains from surface and thermal mismatches during fabrication processes. We set $D = 0.1 \text{ eV} \cdot \text{Å}$ in accordance with our previous work. Finally, after the construction of full Hamiltonian, we apply coordinate transformation $(k_1, k_2, k_3) \rightarrow (k_x, k_y, k_z)$, because of the (112) surface and $[1\bar{1}0]$ edge direction of our real nanoplate sample. $x, y$ and $z$ correspond to longitudinal, transversal, and out-of-plane directions of our device, respectively. The same transformation is applied to $B$ as well.

**Table S1. Model parameters for effective Hamiltonian of $Cd_3As_2$.** [37,62]

| | | | |
|---|---|---|---|
| $E_0$ | -0.0145 eV | $A$ | 0.889 eV·Å |
| $E_1$ | 10.59 eV·Å² | $D$ | 0.1 eV·Å |
| $E_2$ | 11.5 eV·Å² | $g_{1z}$ | 10.0316 |
| $M_0$ | -0.0205 eV | $g_{1p}$ | 11.6211 |
| $M_1$ | 18.77 eV·Å² | $g_{2z}$ | -4.3048 |
| $M_2$ | 13.5 eV·Å² | $g_{2p}$ | 0.5876 |

*Note 11. 2. Slab configuration and subband dispersion*

The introduction of vector potential to the effective Hamiltonian causes problem about commutation between momentum $k$ and spatial coordinate $(x, y, z)$. To be specific, as magnetic field is applied on $x$ direction in our experiments, we choose the vector potential to be $\boldsymbol{A} = (0, -Bz, 0)$, so the translation symmetry on $z$ direction is broken. To solve this situation, we apply the formal method of expanding wavefunctions with plane-wave basis on $z$ direction, i. e. the eigenfunction of the Hamiltonian can be expressed as [63]

$$\psi = \sum_n \psi_n(k_x, k_y) \sin\left[\frac{n\pi}{t}\left(\frac{t}{2} + z\right)\right], \qquad -\frac{t}{2} < z < \frac{t}{2}, \qquad \text{(S9)}$$

with $t$ the thickness of the sample. Such method has been widely adopted to deal with quantum confinement effect on thin slabs of real materials previously. In practice, we set maximum of $n$ to be 50, and focus on the structure of 16 lowest energy subbands (closest to the original Dirac point at $\mu = 0$), which is sufficient to recover key properties of the system. We would like to address that the thickness of our sample (95 nm) exceeds the critical thickness ($t_{c2} = 21$ nm) for a finite topological gap formation, as reported in literature [37], so our model remains gapless when no magnetic field is applied.

*Note 11. 3. Calculation of quantum metric, Berry curvature and corresponding nonlinear Hall conductivities*

The above treatment effectively removes the $z$-dependent physics in our model, reducing it to a 2D-like case, which helps simplify the calculation [6,44,59,64]. The quantum metric and Berry curvature associated with certain band index $n$ can be expressed as $g_{\alpha\beta}^n = \text{Re} \sum_{m\neq n} \mathcal{A}_{nm}^\alpha \mathcal{A}_{mn}^\beta$ and $\Omega_{\alpha\beta}^n = -2\text{Im} \sum_{m\neq n} \mathcal{A}_{nm}^\alpha \mathcal{A}_{mn}^\beta$, with $\mathcal{A}_{nm}^\alpha = \langle n | i \partial_{k_\alpha} | m \rangle$ the interband Berry connection, and $\alpha, \beta = x, y$ the cartesian coordinates in the transport plane. According to Ref. [25], the dipole structure of quantum metric in band $n$, associated with nonlinear Hall transport, is defined as

$$D_{\text{QM}}^n = \int_k \left( v_y^n g_{xx}^n - v_x^n g_{yx}^n \right) \frac{\partial f(\varepsilon_n)}{\partial \varepsilon_n}, \tag{S10}$$

with $v_\alpha^n = \frac{\partial \varepsilon_n}{\hbar \partial k_\alpha}$ the group velocity. In the main text Fig. 3(d), the distribution of the integrand $\left( v_y^n g_{xx}^n - v_x^n g_{yx}^n \right)$ is illustrated for the lowest conduction subband. Here in Fig. S9, we indicate the $\left( v_y^n g_{xx}^n - v_x^n g_{yx}^n \right)$ value upon the (line-cut) band structure by the red and blue color scales. The band edges as well as band crossings give rise to concentration of quantum metric quantities as expected, and the distribution evolves with magnetic field as expected.

The nonlinear Hall conductivity contributed by quantum metric is given by

$$\sigma_{yxx}^{\text{INT}} = 2e^3 \text{Re} \sum_{n,m}^{\varepsilon_n \neq \varepsilon_m} \int_k \left( \frac{v_y^n \langle u_n | i\partial_{k_x} | u_m \rangle \langle u_m | i\partial_{k_x} | u_n \rangle}{\varepsilon_n - \varepsilon_m} \right.$$
$$\left. - \frac{v_x^n \langle u_n | i\partial_{k_y} | u_m \rangle \langle u_m | i\partial_{k_x} | u_n \rangle}{\varepsilon_n - \varepsilon_m} \right) \frac{\partial f(\varepsilon)}{\partial \varepsilon} . \quad \text{(S11)}$$

For the Berry curvature and the associated time-reversal-even nonlinear Hall response, although the topic is not the major focus of current study, it provides side evidences for the physical scenario of Dirac band evolution, and is worth analyzing numerically as well. In Fig. 3(a)-(c) we plot the schematic distribution of Berry curvature associated with each Weyl cone, which is presented by the thickness of red or blue colors on the cone dispersion. Straightforwardly, as a result of their inherent similarity and correlation, the evolution patterns understood for quantum metric – such as extended dipole structure from Weyl cone separation and overall suppression in magnitude due to gap opening – are also shared by Berry curvature. For a more quantitative description, we plot the corresponding distribution of Berry curvature in $k$-space in Fig. S10, where $\Omega_z^n \equiv \Omega_{xy}^n$ is the only nonzero Berry curvature component in 2D case. The net Berry curvature dipole from all bands is defined as

$$D_\alpha = \sum_n \int_k (\partial_{k_\alpha} \Omega_z^n) f(\varepsilon_n), \quad \text{(S12)}$$

and contribute to nonlinear Hall conductivity by [6]

$$\sigma_{yxx}^{\text{BCD}} = \sigma_{xx} \frac{e^3 \tau}{2\hbar^2} D_x = \sigma_{xx} \frac{m^* e}{2\hbar^2 n} D_x. \quad \text{(S13)}$$

This is exactly the equation (S3) which we used for experimental estimation of $D_x$. The dependence of $D_x$ on chemical potential $\mu$ and magnetic field $B$, as well as a comparison between theoretical and experimental results are presented in Fig. S11 and 3(g) in the main text.

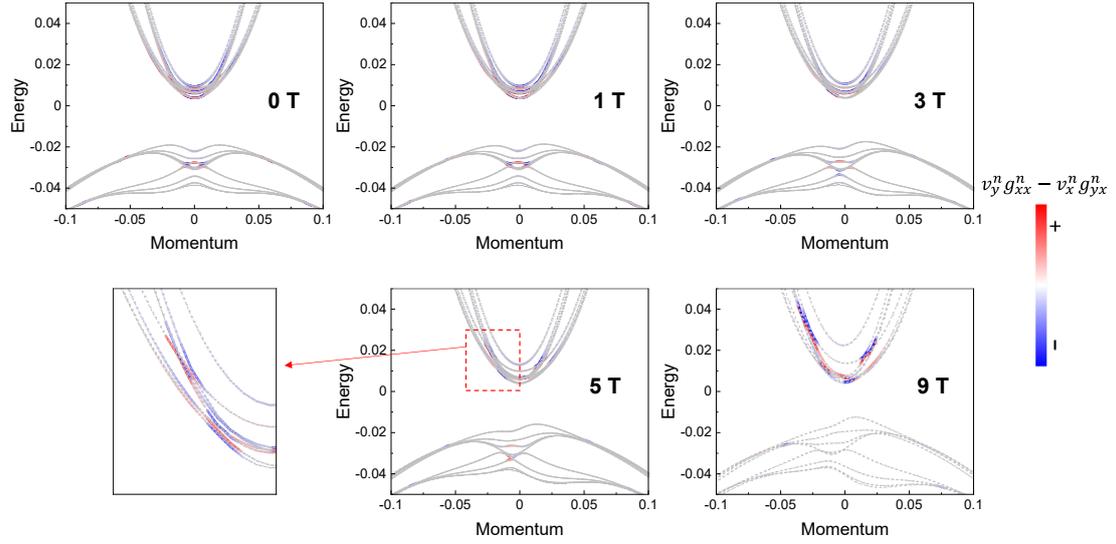

**FIG. S9. Band structures and profiles of $\left(v_y^n g_{xx}^n - v_x^n g_{yx}^n\right)$ along line-cuts in the Brillouin zone, under different magnetic fields.** Red and blue colors indicate the magnitude of the $\left(v_y^n g_{xx}^n - v_x^n g_{yx}^n\right)$ quantity. The units of the axes in each panel are $\text{Å}^{-1}$ and eV, respectively.

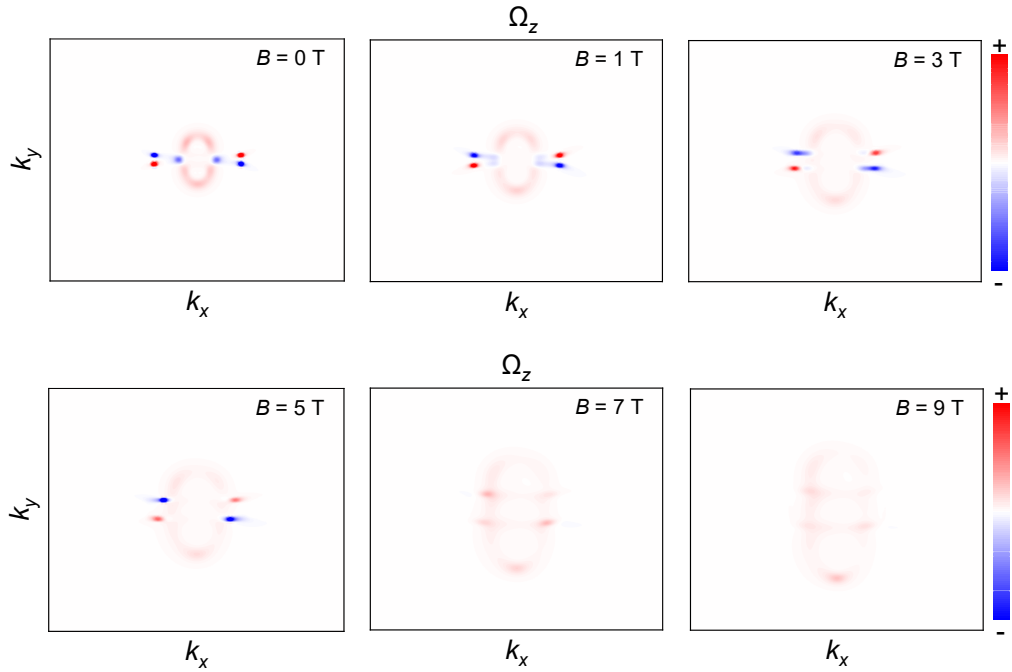

**FIG. S10. Distribution of Berry curvature $\Omega_z$ in the momentum space, under various magnetic fields.** The corresponding value of $B$ is identified in each figure.

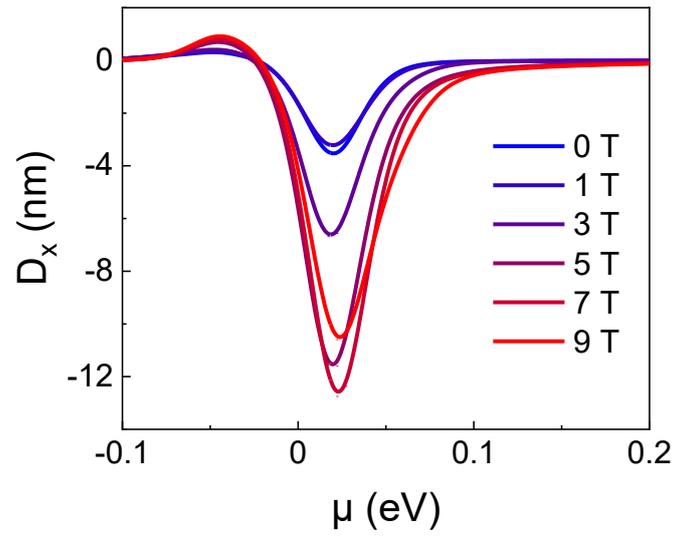

**FIG. S11. Calculated BCD component $D_x$ as a function of chemical potential $\mu$ under various magnetic fields.**

**Note 12. Exclusion of trivial extrinsic mechanisms for nonlinear transport**

*Note 12. 1. Thermoelectric and thermal effects*

In the presence of magnetic field, the nonlinear transport signal may comprise contribution from thermoelectric effects like Nernst effect [65], which lead to a time-reversal-odd nonlinear signal $\sigma_{yxx}^{(o)}$. In our configuration of measurement, to realize a Nernst signal along $y$ direction by magnetic field along $x$ direction, the temperature gradient $\nabla T$ is required to exist along out-of-plane $z$ direction. While this is possible to arise from thermal dissipation through sample-substrate interface, the following analysis evidences that such thermoelectric effect is not responsible for our key observations.

- <u>*Nonlinear and non-monotonic $\sigma_{yxx}^{(o)} - B$ relation.*</u> The variance in magnetoresistance shown in Fig. 1(b) is moderate, especially at the Dirac point $V_g = -10$ V, where longitudinal resistance varies no more than 20% within the range $[0, 9\ \text{T}]$. Thus, under a fixed value of longitudinal bias $V_{xx}$, thermal gradient $\nabla T \sim I^2 R_{xx}$ is almost unchanged across various $B$, and the Nernst signal $\propto B\nabla T$ should exhibit approximately linear dependence on $B$. Nevertheless, our data ubiquitously show non-monotonic $\sigma_{yxx}^{(o)} - B$ relation with certain saturating behavior, which cannot be explained by the Nernst effect.

- <u>*Scaling dependence on $\tau$.*</u> The Nernst signal should be $\tau$-dependent, for thermal gradient $\nabla T \sim I^2 R_{xx}$, and $R_{xx}$ contributes the dependence on $\tau$. Since we mainly focus on the $\tau^0$ term in scaling law, the $\tau$-dependent Nernst term, even if it exists, merely gives correction to scattering-related $\tau^1$ or $\tau^2$ terms, and does not affect our major conclusion about quantum geometry contributions.

- <u>*Temperature dependence of nonlinear Hall signal.*</u> Under a fixed bias voltage $V_{xx}$, the thermal gradient $\nabla T \sim I^2 R_{xx} = \frac{V_{xx}^2}{R_{xx}}$ becomes larger when $R_{xx}$ takes a smaller value. According to Fig. 2(b) in the main text, our device exhibits a semiconductor-like $R_{xx} - T$ relation, with smaller $R_{xx}$ at higher temperature. Thus, it is expected that Nernst signal is enhanced at high temperature due to a small $R_{xx}$. This contradicts with our observation in Fig. 2(a), where time-reversal-odd nonlinear signal mainly shows a decreasing trend as temperature increases.

Apart from thermoelectric effects, the Joule effect itself may also induce global heating of the sample. Such side effect can also be safely excluded by our measurements due to following reasons. First, the nonlinear signal induced by thermal resistance variation should in principle show cubic dependence on bias current, as $\delta V_{xx} = I\delta R_{xx} \approx I\frac{\partial R_{xx}}{\partial T}\delta T \sim I\frac{\partial R_{xx}}{\partial T}(I^2 R_{xx}) \propto I^3$, leading to third-order response rather than second-order [66]. On the other hand, the resistance variation should also influence the linear longitudinal response, causing a deviation from perfect linear $V_{xx} - I$ relation. As shown in Fig. S12, linear $V_{xx} - I$ relation can be well-observed under various magnetic fields at the Dirac point, showing no obvious deviation when the second-order Hall signal is magnificent [see the highlighted green line in Fig. S12, which corresponds to $B = 3$ T where nonlinear Hall signal is maximized; see also Fig. 1(f)].

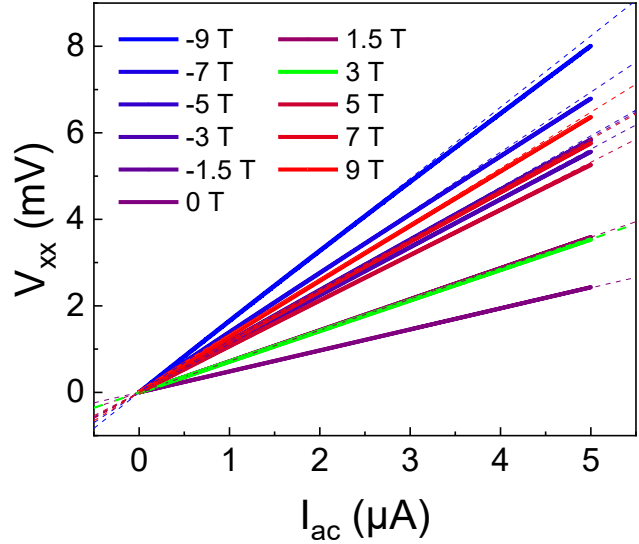

**FIG. S12. Linear $V_{xx} - I$ relation measured at $T = 2$ K, $V_g = -10$ V, under various magnetic fields.** Dashed lines are linear fitting to data within $0 < I_{ac} < 1$ μA. Data associated to $B = 3$ T is highlighted in green.

*Note 12. 2. Longitudinal and transversal signal coupling*

It is common for longitudinal and transversal electric signals to couple and mix in real devices. Either intrinsic factors like resistance anisotropy, or external ones like misalignment from micro-fabrication processes can lead to such consequence. From following aspects, we show that the nonlinear Hall effect in our measurement is a spontaneous transport behavior, rather than a component from longitudinal transport.

Figure S13 plots together the $R_{xx}$ and $R_{xy}$ transfer curves at $T = 2$ K. The $R_{xy} - V_g$ curve clearly resembles the shape of $R_{xx} - V_g$ curve, yet an order smaller in magnitude, suggesting the $R_{xy}$ signal comes from $R_{xx}$ coupling with a coupling ratio $R_{xy} \approx -\frac{1}{10} R_{xx}$. We also measured the nonlinear longitudinal signal $V_{xx}^{2\omega}$ at the Dirac point under various magnetic fields, as shown in Fig. S14(a), simultaneously obtained with nonlinear Hall data in Fig. 1(e). The corresponding nonlinear ratio $\frac{V_{xx}^{2\omega}}{(V_{xx})^2}$ as a function of $B$ is summarized in Fig. S14(b). The time-reversal-odd and even parts of $\frac{V_{xx}^{2\omega}}{(V_{xx})^2}$ are separated in the same manner as $\frac{V_{xy}^{2\omega}}{(V_{xx})^2}$. Interestingly, these components show the similar non-monotonic dependence on magnetic field like the Hall components. It is evident that $\frac{V_{xx}^{2\omega}}{(V_{xx})^2}$ components are of the same order as (or even smaller than) $\frac{V_{xy}^{2\omega}}{(V_{xx})^2}$ components, and the corresponding components ($\frac{V_{xx-\text{even}}^{2\omega}}{(V_{xx})^2}$ and $\frac{V_{xy-\text{even}}^{2\omega}}{(V_{xx})^2}$ for instance) have the same sign. These facts contradict with the $-\frac{1}{10}$ coupling factor obtained from first-harmonic responses, remarking the nonlinear Hall effect to be spontaneously generated instead of coupled.

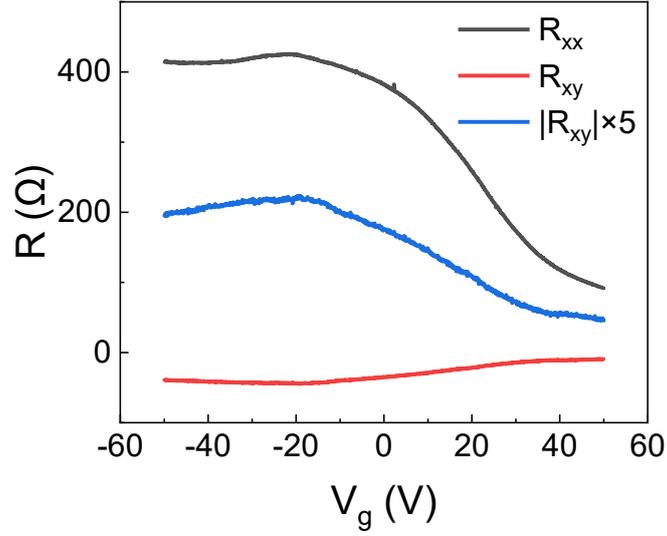

**FIG. S13. Transfer curve of both longitudinal (black) and Hall (red) resistance.** Blue curve helps identify the order of difference between $R_{xx}$ and $R_{xy}$.

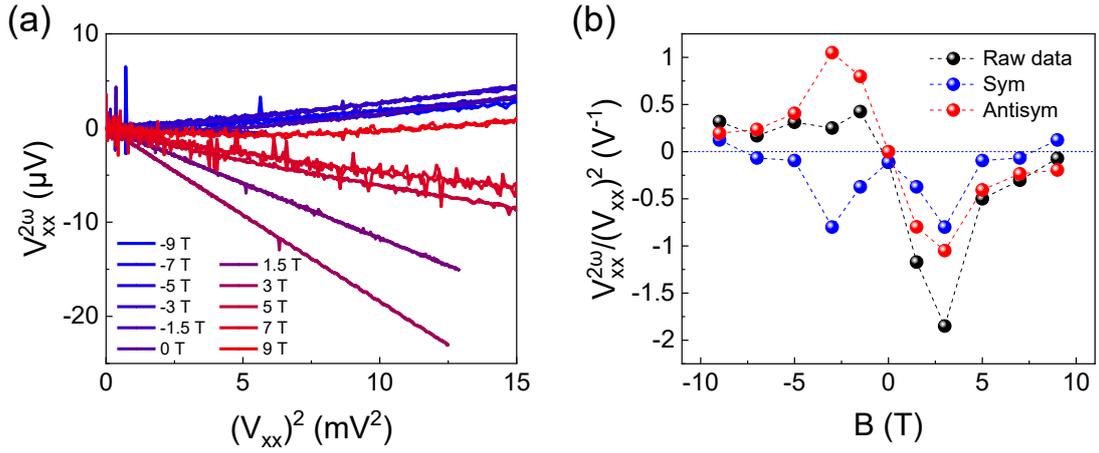

**FIG. S14. Exclusion of longitudinal nonlinear signals as a source to the nonlinear Hall effect.**

(a) Raw data of nonlinear longitudinal signal $V_{xx}^{2\omega}$ under various magnetic field.

(b) Raw data and (anti-)symmetrization of the longitudinal nonlinear signal ratio $\frac{V_{xx}^{2\omega}}{(V_{xx})^2}$. All data are measured at the Dirac point $V_g = -10$ V under $T = 2$ K.

*Note 12. 3. Rectification from diode effect at contacts*

Poor contact condition is also possible to emerge during the device fabrication process, if interface is not properly cleaned before deposition. The corresponding

Schottky barrier may also induce nonlinear $V - I$ relations. However, the highly linear $V_{xx} - I$ curves shown in Fig. S12 guarantee a good ohmic contact of our device. Also, the nonlinear transport from diode effect is not likely to show dependence on magnetic field, especially a $B$-odd one. Therefore, we conclude that diode effect cannot make considerable contribution to the nonlinear Hall effect.

*Note 12. 4. Capacitive circuit coupling*

Parasitic capacitance may exist in external circuits and cause certain nonlinearity. To exclude such interference, we carry out frequency-dependent nonlinear Hall measurements. Figure S15 shows nonlinear Hall voltage $V_{xy}^{2\omega}$ versus bias current $I_{ac}$, measured at $T = 2$ K, $B = 3$ T, $V_g = -10$ V, with driving frequency ranging from 17.777 Hz to 1777.7 Hz. No clear frequency dependence is observed in our measurement, proving negligible influence from capacitive circuits.

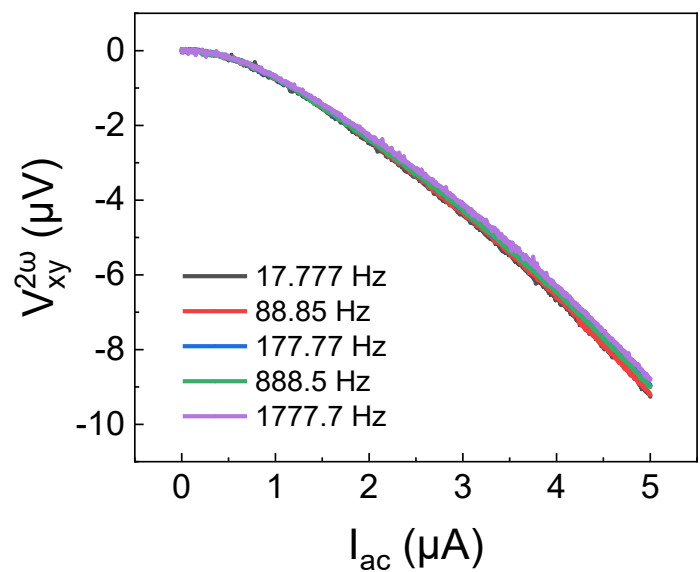

**FIG. S15. Frequency dependence of the nonlinear Hall effect measured at $T = 2$ K, $V_g = -10$ V and $B = 3$ T.**